\documentstyle[aps,epsfig]{revtex}
\begin{document}
\draft
\preprint{\vbox{\hbox{IFT--P.0XX/98}\vspace{-0.3cm}
                \hbox{hep-ph/9811374} }}
%
\twocolumn[\hsize\textwidth\columnwidth\hsize\csname
@twocolumnfalse\endcsname
\title{Three Jet Events and New Strong Couplings at LEP and NLC}
\author{A.\ Belyaev $^{1,2}$, F.\ de Campos $^3$, S.\ F.\ Novaes $^1$, 
and R.\ Rosenfeld $^1$}
\address{
$^1$ Instituto de F\'{\i}sica Te\'orica, Universidade  Estadual Paulista,
\\  
     Rua Pamplona 145, 01405--900, S\~ao Paulo, Brazil. \\
$^2$ Skobeltsyn Institute of Nuclear Physics, Moscow State University \\
     119899 Moscow, Russian Federation. \\
$^3$ Depto.\ de F\'{\i}sica e Qu\'{\i}mica, Universidade Estadual Paulista
\\
     Av.\ Dr.\ Ariberto Pereira da Cunha 333, 12500 Guaratinguet\'a,
Brazil}
\date{\today}
\maketitle
\widetext
\begin{abstract}
We study the effects of new dimension--$6$ operators,  resulting
from a general $SU(3)_C \otimes SU(2)_L \otimes U(1)_Y$ invariant
effective Lagrangian, on three jet production at LEP and at the
Next Linear Collider. Contributions to the total event rate and
to some event shape variables are analysed in order to establish
bounds on these operators.
\end{abstract}
\vskip2pc]
\narrowtext

Quantum Chromodynamics (QCD), an important part of the Standard
Model (SM),  has been tested in the perturbative regime to a high
degree of precision \cite{QCD}. However, the possible existence
of new physics beyond the Standard Model, involving heavy colored
particles, may give rise to small effects in QCD phenomenology at
present and future colliders. Certainly, one of the main goals of
the future generation of colliders will be to scrutinize the
several competitive models describing the physics at high
energies. 

On the phenomenological side, instead of concentrating on a
specific model, it is in general quite instructive to make a
model independent analysis of the indirect effects that an
unknown high--energy theory can have at the present energy scale.
This can be accomplished by the effective Lagrangian approach
\cite{effective}. After the heavy fields of the high--energy
theory have been integrated out, their low--energy consequences
can  be represented by a series of local $SU(3)_C \otimes SU(2)_L
\otimes U(1)_Y$ invariant operators built from the light Standard
Model fields:
\begin{equation}
{\cal L}_{\text{eff}} = {\cal L}_{0} + \sum_{n = 1 \cdots}
           \frac{f_{(n + 4)}}{\Lambda^{n}} \, { O}_{(n + 4)} 
\label{lag:eff}
\end{equation}
where ${\cal L}_{\text{SM}}$ is the Standard Model Lagrangian,
$\Lambda$ is the characteristic scale of the new physics and 
${O}_{(n+4)}$ are the local operators of dimension $(n + 4)$.
Different scenarios can generate the same kind of operator but
with distinct effective couplings $f_{n + 4}$ making possible, at
least in principle, to point out a specific model for the new
physics. 

The classification of the operators ${O}_{(n+4)}$ have been
first done in Ref.\ \cite{buc:wyl} and since then the
phenomenological implications have been studied in the bosonic
sector of the SM \cite{eff:bos}, and for the third--family quarks
\cite{eff:top}. There have also been many studies of the
so--called purely gluonic operators \cite{pure:gluon} where the
high dimension operators ${O}_{(n+4)}$ involves only the gluon
field and modify the non--abelian three  and four vertex. 

Nevertheless, effective operators involving gluons and light
quarks (and possibly the Higgs fields) can also give rise to some
measurable effects in QCD processes at the present colliders.
These new couplings can be generated via loops of colored objects
belonging to the underlying theory \cite{loops}. In this letter
we search for possible signals of the existence of these new
couplings in three jet events at $e^+e^-$ colliders. We analyze
the total event rate for different values of the jet resolution
variable ($y_{\text{cut}}$). Event shape observables in
$e^+ e^-$ colliders are important to test QCD and have been
studied at PETRA \cite{petra}, LEP1 \cite{lep1} and LEP2
\cite{lep2} energies. Therefore, we also explore the differences
in the event shape distributions due to the anomalous
contribution in order to establish bounds on the coefficient of
the dimension 6 operators that alter the $qqg$ interaction. 

In order to study the possible deviation from the Standard Model
predictions for the couplings involving quarks and gluons, we
start by writing the most general dimension--6 effective
Lagrangian requiring the $SU(3)_C \otimes SU(2)_L \otimes U(1)_Y$
invariance of the new operators. We assume that there are no additional
new fields and we construct these operators taking into account
only the usual light particles, {\it i.e.\/} gauge bosons and
quarks.  Furthermore, we do not consider here the operators that
modify the couplings of the gauge bosons with fermions since they
are strongly constrained by the LEP1 measurements at the $Z^0$
pole. Therefore the new Lagrangian can be written as \cite{buc:wyl},
\begin{equation}
{\cal L}_{2} = \frac{1}{\Lambda^2} \, \sum_i \, A_i \, {\cal O}_i \; , 
\label{lagrangian}
\end{equation}
where $A_i$ are constants and the dimension--6 operators ${\cal
O}_i$ can either involve just quarks and vector bosons or may
contain also the Higgs field. In the first case, we have, 
\begin{mathletters}
\label{q:g}
\begin{eqnarray}
{\cal O}_{Qg} &=& i \; \left( \bar{Q} \lambda^a \gamma^\mu 
{\cal D}^\nu Q \right) \; G_{\mu\nu}^a  + \text{h.c.} \; , \label{Qg} \\
{\cal O}_{Ug} &=& i \; \left( \bar{U} \lambda^a \gamma^\mu 
{\cal D}^\nu U \right) \; G_{\mu\nu}^a  + \text{h.c.} \; , \label{Ug} \\
{\cal O}_{Dg} &=& i \; \left( \bar{D} \lambda^a \gamma^\mu 
{\cal D}^\nu D \right) \; G_{\mu\nu}^a  + \text{h.c.}\; , \label{Dg}  
\end{eqnarray}
\end{mathletters}
where $Q$ are the left--handed quark doublets while $U$ and $D$
are the right--handed quark singlets. $G_{\mu\nu}^a =
\partial_\mu G_\nu^a - \partial_\nu G_\mu^a + g_s f^{abc} G_\mu^b
G_\nu^c$ is the usual $SU(3)_C$ strength tensors and ${\cal D}_\mu
= \partial_\mu - i g_s (\lambda^a/2) G_\mu^a - i g (\tau^i/2)
W^i_\mu - i g^\prime Y B_\mu$ is the $SU(3)_C \otimes SU(2)_L \otimes
U(1)_Y$ covariant derivative of the quarks. The operator (\ref{Qg})
gives rise to a new $qqg$ vertex involving left--handed up and
down quarks while (\ref{Ug}) and (\ref{Dg}) operators involve
right--handed up and down quarks respectively. Therefore, if we
assume that the quark--gluon coupling is blind to the quark
flavors, {\it i.e.\/} universal, and that the new physics affects
left and right--handed quarks in the same way, we should require
that $A_{Qg} = A_{Ug} = A_{Dg} \equiv A_{qg}$. We should  point
out that the new interactions (\ref{q:g}) also generate  new
couplings involving weak--vector bosons ($V$), like  $qqgV$ and
$qqggV$, and also vertex with quarks and two and three gluons.

The operators that involves also the Higgs field doublet ($\phi$)
can be written as, 
\begin{mathletters}
\label{q:g:h}
\begin{eqnarray}
{\cal O}_{Ug\phi} &=& \left( \bar{Q} \sigma^{\mu\nu} \lambda^a U \right) 
\tilde{\phi} \; G_{\mu\nu}^a + \text{h.c.} \; , \label{ugh} \\ 
{\cal O}_{Dg\phi} &=& \left( \bar{Q} \sigma^{\mu\nu} \lambda^a D \right) 
\phi \; G_{\mu\nu}^a + \text{h.c.} \; ,  \label{dgh} 
\end{eqnarray}
\end{mathletters} 
\noindent
where $\sigma^{\mu\nu} = (i/2) [\gamma^\mu, \gamma^\nu]$. When
$\phi$ is replaced by its vacuum expectation value, the operators
(\ref{q:g:h}) generate new $qqg$, and $qqgg$ interactions, for
$q= u, \; d$ quarks. In order to guarantee the universality also
in the magnetic type $qqg$ coupling, we should assume that
$A_{Ug\phi} = A_{Dg\phi} \equiv A_{qg\phi}$.

Therefore, we end up with the following new Lagrangians,
\begin{eqnarray}
{\cal L}_{qg} &=&  \frac{2 A_{qg}}{\Lambda^2} 
\Biggl\{ 
\frac{i}{2} \sum_{q} \left[ \bar{q} \lambda^a \gamma^\mu (\partial^\nu q)
- (\partial^\nu \bar{q}) \lambda^a \gamma^\mu  q  \right] 
\nonumber \\
&+& \frac{g_s}{2}\sum_{q} \left( \bar{q} \{ \lambda^a,
\lambda^b\} \gamma^\mu q \right) G_{\nu}^b 
\nonumber \\
&+& e \sum_{q} Q_q \left(\bar{q} \lambda^a \gamma^\mu q \right) A^\nu
\nonumber \\
&+& \frac{e}{s_W c_W } 
\sum_{q} \left[ \bar{q} \lambda^a \gamma^\mu (g_V^q + g_A^q
\gamma_5) q \right] Z^\nu 
\nonumber \\
&+& \frac{e}{2 \; \sqrt{2} s_W}
\sum_{u,d} \Bigl[ \bar{u} \lambda^a \gamma^\mu (1- \gamma_5) d \;
W^{+ \; \nu} 
\nonumber \\
&+& \bar{d} \lambda^a \gamma^\mu (1- \gamma_5) u \;  W^{- \; \nu}
\Bigr]   \Biggr\} G_{\mu\nu}^a 
 \; , \label{qg} 
\end{eqnarray}
where the summation is made over all the quark flavors $q$ and
over up and down quarks ($u$, $d$). $g_V^q = T_3^q/2 - Q_q
s_W^2$ and $g_A^q = - T_3^q/2$ with $s_W$ being the sine of the
Weinberg angle, $T_3^q$ and $Q_q$ being the quark weak isospin
and electric charge respectively, and 
\begin{equation}
{\cal L}_{qg\phi} = \frac{A_{qg\phi}}{\Lambda^2} \frac{(v + H)}{\sqrt{2}} 
\sum_{q} \left( \bar{q} \, \sigma^{\mu\nu} \, \lambda^a \, q \right)
G_{\mu\nu}^a 
\; , \label{qgh} 
\end{equation}

We shall start by studying the sensitivity to these new higher
dimensional operators at LEP1, which has accumulated a large data
sample of three jet events. This analysis was performed by
including the new couplings  generated by the higher dimensional
operators into the package CompHEP \cite{comp}. We found that
there is no contribution of the operators ${\cal O}_{qg}$ when
the gluon is on--shell, like in the process $e^+ e^- \rightarrow
q \bar{q} g$. Furthermore, for the contributions generated by the
${\cal L}_{qg\phi}$ Lagrangian there is no interference with the
SM amplitudes.

In order to compare with LEP1 data, we used the OPAL
Collaboration \cite{OPAL1} best fit values for the relevant
energy scale ($Q^2 = (6.4 \; \mbox{GeV})^2$) and for the QCD
scale ($\Lambda_{\text{QCD}} =  147$ MeV). In this way we
effectively minimize the uncertainty due to next--to--leading
order corrections. We employed the JADE jet algorithm
\cite{jade:alg} by requiring that the three final state partons
obey:
\begin{equation}
y_{ij} \equiv \frac{M_{ij}^2}{s} > y_{\text{cut}}
\end{equation}
for any pair of final state partons, where $M_{ij}$ is the
invariant mass of the $(i,j)$ pair and $y_{\text{cut}}$ is a
parameter that determines the jet separation criteria used
experimentally.  We have checked that our result do not change in
a significant way if we consider the Durham \cite{dur:alg} or
Cambridge \cite{cam:alg} jet algorithms where $M_{ij} = 2 \;
\text{min}(E_i^2, E_j^2) (1 - \cos\theta_{ij})$. 

\begin{figure}[htb]
\vskip -2cm
\begin{center}
\mbox{\epsfig{file=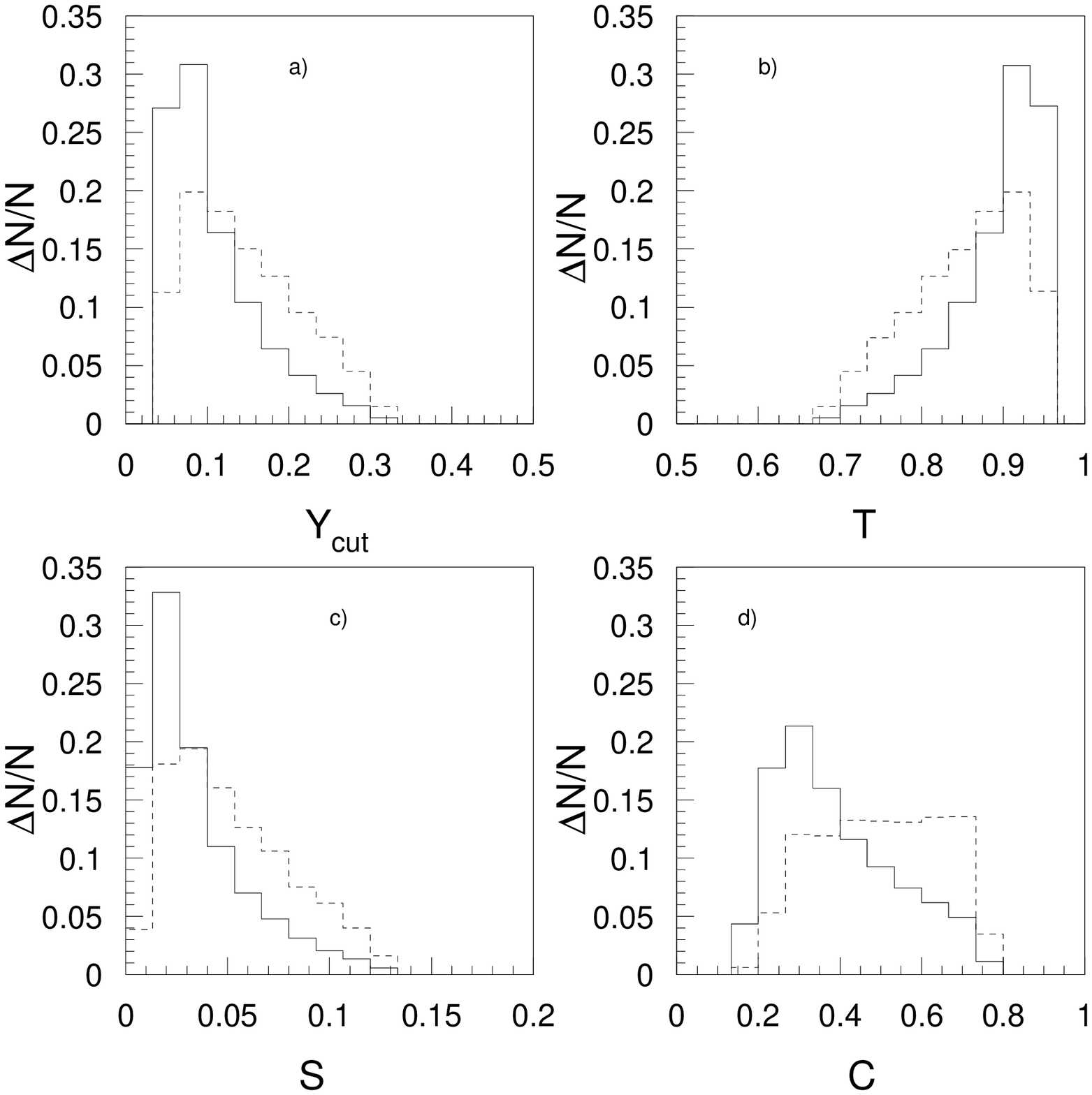,width=0.95\textwidth}}
\end{center}
\caption{Relative production rate of three jet events as a
function of $y_{\text{cut}}$ (a), and the normalized
distributions for the event shape variables: thrust (b),
spherocity (c) and C--parameter (d), for  SM (solid line) and
pure anomalous interactions (dashed line). In all cases we have
considered $y_{\text{cut}}^{\text{min}} = 0.05$.}
\label{fig:1}
\end{figure}

In our analysis, we assumed $y_{\text{cut}}^{\text{min}} = 0.05$
and we analyzed, besides the relative production rate of three
jet events as a function of $y_{\text{cut}}$, different event
shape distributions, like thrust ($T$) \cite{t}
\begin{eqnarray}
T & = & \mbox{max}_n \frac{ \sum_{i} | p_i \cdot n | }{ \sum_{i} 
| p_i | } \; ,
\end{eqnarray}
spherocity ($S$) \cite{s},
\begin{eqnarray}
S & = & \left( \frac{4}{\pi}\right)^2 \mbox{min}_n \left( 
\frac{ \sum_i | p_i \times n | }{ \sum_i | p_i | } \right)^2  \; ,
\end{eqnarray}
and the $C$--variable \cite{c},
\begin{eqnarray}
C & = & \frac{3}{2} \frac{ \sum_{i,j} \left[ | p_i | | p_j |  - 
(p_i \cdot p_j)^2/
| p_i | | p_j | \right] }{ (\sum_i | p_i | )^2}  \; .
\end{eqnarray}
In order to illustrate the shape of these distributions,  we
present in Fig.\ \ref{fig:1},  our results for $y_{\text{cut}}$,
$S$, $T$ and $C$ normalized distributions for the Standard
Model and for the pure anomalous case.  

We performed a $\chi^2$ analysis for the various distributions to
estimate the sensitivity of the three jet events to the anomalous
parameter. We have taken into account the statistical errors and
the overall normalization uncertainty of the QCD prediction. We
consider,
\[
\chi^2 = \sum_{i} \frac{[N_i - f N_i^{\text{SM}}]^2}{f N_i^{\text{SM}}}=
\sum_{i} \frac{[N_i^{\text{ANO}}+(1-f)N_i^{SM}]^2}{f N_i^{\text{SM}}}
\]
where, $N_i$ and $N_i^{\text{SM}}$ are the numbers of events in
the  $i$th histogram bin in the presence of anomalous coupling
and for the pure standard case, while $N_i^{\text{ANO}} = N_i -
N_i^{\text{SM}}$ and $f$ is a normalization parameter which
parametrizes the changes in the overall QCD normalization. We
have minimized $\chi^2$ with respect to $f$ in order to restrict
$\chi^2$ sensitivity only to the shape difference between
anomalous and the Standard Model scenarios. In our analysis we
assumed that the dominant errors are statistical and
fragmentation and detector effects could be ignored. 

Assuming a total luminosity of $220 \; \mbox{pb}^{-1}$ \cite{lepwg}
we derived the following $95 \%$ CL. bounds from the various
distributions,
\begin{eqnarray}
\frac{A_{qg\phi}}{\Lambda^2} & < & 16.3 \; \mbox{TeV}^{-2}, 
\; \mbox{from $y_{\text{cut}}$} \\
\frac{A_{qg\phi}}{\Lambda^2} & < & 14.2 \; \mbox{TeV}^{-2}, 
\; \mbox{from thrust} \\
\frac{A_{qg\phi}}{\Lambda^2} & < & 16.0  \; \mbox{TeV}^{-2}, 
\; \mbox{from spherocity}\\
\frac{A_{qg\phi}}{\Lambda^2} & < & 16.1  \; \mbox{TeV}^{-2}, 
\; \mbox{from C--parameter}
\end{eqnarray}
It is important to notice that these bounds decrease by only
$\sim 15 \%$ if we assumed the value $Q^2 = M_Z^2$ for the QCD
energy scale instead of the OPAL best fit value. In fact there is
not a very good improvement on the bounds obtained from the event
shape distribution when compared with the ones coming from the
total yield: the thrust gives a slightly better bound. Therefore,
we are able to establish the bound of $\Lambda \gtrsim 270$ GeV,
for $A_{qg\phi} = 1$, while for $A_{qg\phi} = 4 \pi$, $\Lambda$
should be larger than $1$ TeV. 

We have also repeated the same analysis for LEP2 energies
($\sqrt{s} \simeq 200$ GeV) and 200 pb$^{-1}$ of data and also
for the Next Linear Collider (NLC) assuming a center--of--mass
energy of $\sqrt{s} = 500$ GeV and $\sqrt{s} = 1$ TeV with an
integrated luminosity of 100 fb$^{-1}$. 

At LEP2, since we are far from the $Z^0$ peak, we get a weaker
bound on the scale of $\Lambda \gtrsim 140$ GeV ($A_{qg\phi} =
1$). However, at NLC with higher energies and luminosities, we
can improve our bounds. The relative contribution from anomalous
interaction grows with the energy while the SM cross section
falls down. At $\sqrt{s} = 500$ GeV, NLC is able to establish the
limit of $\Lambda \gtrsim  390$ GeV, for $A_{qg\phi} = 1$. When
we further increase the energy to $\sqrt{s} = 1$ TeV the bound
becomes: $\Lambda \gtrsim 480$ GeV, for $A_{qg\phi} = 1$. 

In this letter, we have shown how the study of three jet
production at an $e^+e^-$ collider can provide an important test
of $qqg$. In particular, we derived for the first time direct
bounds on the anomalous couplings involving light quarks, gluons and
the Higgs boson. These direct bounds are obtained from the study
of the total cross section and also from the event shape
variables distributions. Similar operators to the ones studied
here have been recently constrained by Gounaris, Papadamou and
Renard \cite{eff:top} using unitarity arguments. However, these
indirect bounds are important only for operators involving the
top quark, and hence cannot be applied to the operators discussed
in the present work.  In conclusion, the comparison of anomalous
contribution to the $qqg$ vertex with the QCD predictions can be
quite sensitive to new physics effect.

{\it Acknowledgments:} This work was supported by Conselho
Nacional de Desenvolvimento Cient\'{\i}fico e Tecnol\'ogico
(CNPq), by Funda\c{c}\~ao de Amparo \`a Pesquisa do Estado de
S\~ao Paulo (FAPESP), and by Programa de Apoio a N\'ucleos de
Excel\^encia (PRONEX).


\newpage

~

\newpage
\widetext

\end{document}